\documentclass{aa} 
\usepackage{graphicx}
\usepackage{hyperref}
\usepackage{natbib}                 
\bibpunct{(}{)}{;}{a}{}{,}          

\usepackage{amsmath}                
\usepackage{psfrag}                 
\usepackage{txfonts}                

\begin{document}
  \title{A synchrotron self-Compton scenario for the very high energy $\gamma$-ray emission of the radiogalaxy M\,87}
  
  \subtitle{Unifying the TeV emission of blazars and other AGNs ?}
  
  \author{J.-P.~Lenain
    \inst{1}
    \and
    C.~Boisson\inst{1}
    \and
    H.~Sol\inst{1}
    \and
    K.~Katarzy{\'n}ski\inst{2}
  }
  
  \offprints{J.-P.~Lenain\\
    \email{\href{mailto:jean-philippe.lenain@obspm.fr}{jean-philippe.lenain@obspm.fr}}}
  
  \institute{LUTH, Observatoire de Paris, CNRS, Universit{\'e} Paris Diderot; 5 Place Jules Janssen, 92190 Meudon, France
    \and
    Toru\'n Centre for Astronomy, Nicolaus Copernicus University, ul. Gagarina 11, 87100 Toru\'n, Poland
  }

  \date{Received 1 June 2007 / Accepted 12 October 2007}
  
  \abstract
      {M\,87 is the first extragalactic source detected in the TeV range that is not a blazar. With the increasing performances of ground-based \v{C}erenkov telescopes, we can now probe the variability in the $\gamma$-ray flux at small timescales, thus putting strong constraints on the size of the emitting zone. The large scale jet of M\,87 is misaligned with respect to the line of sight. A modification of standard emission models of TeV blazars appears necessary to account for the $\gamma$-ray observations despite this misalignment.
      }
      {We explain TeV $\gamma$-ray spectra and fast variability of M\,87 by invoking an emission zone close to the central supermassive black hole, which is filled with several plasma blobs moving in the large opening angle of the jet formation zone.
      }
      {We develop a new multi-blob synchrotron self-Compton (SSC) model with emitting blobs set on a cap beyond the Alfv{\'e}n surface in the jet, at a distance of $\sim$100\,$r_g$ from the central engine to interpret the high energies inferred by new TeV observations. We present a SSC model that is explicitly adapted to deal with large viewing angles and moderate values of the Lorentz factor inferred from (general relativistic) magnetohydrodynamic models of jet formation.
      }
      {This scenario can account for the recent $\gamma$-ray observations of M\,87 made by the High Energy Stereoscopic System (H.E.S.S.) telescope array. We find individual blob radii of about $10^{14}$\,cm, which is compatible with the variability on timescales of days recently reported by the H.E.S.S. collaboration and is of the order of the black hole gravitational radius. Predictions of the very high energy emission for three other sources with extended optical or X-ray jet which could be misaligned blazars still with moderate beaming are presented for one Seyfert 2 radiogalaxy, namely Cen\,A, one peculiar BL\,Lac, PKS\,0521$-$36, and one quasar, 3C\,273.
      }
      {}
      
      
      \titlerunning{A SSC scenario for the VHE $\gamma$-ray emission of M\,87}
      
      \maketitle

      %
      \section{Introduction}
      
      M\,87 is a well-known nearby giant elliptical galaxy \citep[z=0.00436,][]{2000MNRAS.313..469S} close to the center of the Virgo cluster, which shows a multi-spectral jet, signature of an active galactic nucleus (AGN). Its jet is one of the best known, at all scales, thanks to its nearby location and its strong synchrotron radiation in the optical band. M\,87 is classified as FR\,I based on its radio morphology. \citet{2002ApJ...568..133W} observed the jet with {\it Chandra} on July 29 and 30, 2000 and detected it up to a distance of $\sim$21$\arcsec$ from the core in the X-ray band, which implies that the jet is not as strongly aligned along the line of sight \citep[see also][]{1996MNRAS.283..873R} as in the case of blazars.
     
      At radio wavelengths, an impressive jet, which extends up to a few tens of kiloparsecs, can be seen. The central engine is thought to be a supermassive black hole (SMBH) with a mass of $M_\mathrm{BH} \sim 3 \times 10^9 M_{\sun}$ \citep{1997ApJ...489..579M}. The scale length is thus $r_g = G M_\mathrm{BH} / c^2 \sim 4.5 \times 10^{14}$\,cm $\sim 1.4 \times 10^{-4}$\,pc. Using the {\it Hubble Space Telescope} ({\it HST}), \citet{1999ApJ...520..621B} observed superluminal apparent motions of about $4c$--$6c$ beyond 400\,pc for the internal knots, between 1994 and 1998, thus confirming that the jet is relativistic. They conclude that the jet is oriented within $19 \degr$ of the line of sight.
      
      Due to the presence of a SMBH in the core and the presence of the jet, M\,87 was deemed an interesting candidate for TeV emission. \citet{2004ApJ...610..156L} reported an upper limit observed with Whipple in 2000 and 2001, simultaneously with X-ray flares observed by {\it RXTE}. HEGRA observed M\,87 in 1998 and 1999 for a total exposure of 77\,h after data quality selection \citep{2003A&A...403L...1A,2004NewAR..48..407B}. A 4.1$\sigma$ significance was recorded and an integrated flux ($E>730$\,GeV) of 3.3\% Crab was measured.
            
      Recently, \citet{2006Sci...314.1424A} have observed M\,87 with the High Energy Stereoscopic System (H.E.S.S.)\footnote{\url{http://www.mpi-hd.mpg.de/hfm/HESS/}} between 2003 and 2006 in 89\,h live-time with a $13\sigma$ detection and discovered variations on timescales of about 2 days, 10 times faster than that observed in any other waveband. This shows that the emission region is very compact, with a dimension of the order of a few Schwarzschild radii. These observations, thus confirming the detection by HEGRA \citep[][]{2005astro.ph..4395B}, are particularly interesting since M\,87 is the first non-BL\,Lac extragalactic object ever observed at TeV energy. Radio-loud galaxies contain AGNs with jets like blazars, but the jet emission is less strongly boosted due to larger viewing angles between the jet and the observer's line of sight. It is therefore a challenge for standard models of TeV blazars to explain the very high energy (VHE) emission of M\,87.
      
      In this paper, we present a modified synchrotron self-Compton (SSC) scenario to explain the VHE emission of M\,87. Classic SSC models \citep[e.g.][]{1979A&A....76..306G,1996ApJ...463..555I,1996ApJ...461..657B,1999MNRAS.306..551C,2001A&A...367..809K} are applied to blazars, which are beamed sources, and cannot account for the observations of radiogalaxies like M\,87. Our goal is to further develop one of these models to reconcile beamed and unbeamed sources in the same framework of models. Such propositions for unification of AGNs have already been studied considering orientation effects \citep[e.g.][]{1993ARA&A..31..473A,1995PASP..107..803U}, or radio/X-ray power among BL\,Lac objects, flat-spectrum radio-loud quasars (FSRQs) and FR\,Is \citep[e.g.][]{1998MNRAS.299..433F,1998MNRAS.301..451G,2000MNRAS.318..493C}.

      A short description of the leptonic blob-in-jet model for TeV blazars is found in Sect.~\ref{sec:bij}, and its development and application to M\,87 are described in Sect.~\ref{sec:multi_blobs} and Sect.~\ref{sec:M87}. In the framework of misaligned BL\,Lac-like objects, we then try to predict VHE fluxes for objects with optical/X-ray extended jets in Sect.~\ref{sec:predictions}. Implications on unification schemes of AGNs are discussed in Sect.~\ref{sec:unification}.

      
      \section{``Blob-in-jet'' leptonic SSC model
	\label{sec:bij}}
      
      We intend to model the multiwavelength spectrum of M\,87 in the framework of a quasi-homogeneous SSC scenario, successfully used to account for the overall emission of blazars, such as \object{Mrk 501} and \object{Mrk 421}. Our model relies on the basic scenario presented in \citet[][and references therein]{2001A&A...367..809K,2003A&A...410..101K} who give the details of the computation of the radiative transfer and emission by SSC processes in a single spherical blob of plasma moving at relativistic speed along the jet axis. The blob, immersed in a uniform magnetic field, is assumed to be located inside the jet, close to the central engine. An inhomogeneous conical extended jet model is also used to explain the emission from radio to ultraviolet wavelengths \citep[see Sect.~2.2 in][for more details]{2001A&A...367..809K}. The absorption by the infrared extragalactic background light at VHE is taken into account and modeled using the estimations as described in \citet{2006ApJ...648..774S} and references therein. Here we model only nearby active galactic nuclei (AGNs) and hence this effect can be neglected. The blob-in-jet model is particularly well adapted to the description of blazars, for which the jet is very close to the line of sight. In the following, the assumed cosmology is $H_0 = 70\ $km\,s$^{-1}$\,Mpc$^{-1}$ for an Einstein-de~Sitter universe, with $\Omega_\Lambda=0.7$ and $\Omega_m=0.3$.

      We assume that the population of electrons, which is responsible for the non-thermal emission in leptonic models, has a number density that can be described by a broken power-law:

      \begin{equation}
      N_e(\gamma)=
      \begin{cases}
	K_1 \gamma^{-n_1} & \gamma_\mathrm{min} \leqslant \gamma \leqslant \gamma_\mathrm{br} \\
	K_2 \gamma^{-n_2} & \gamma_\mathrm{br} \leqslant \gamma \leqslant \gamma_c
      \end{cases}
      \quad [\mathrm{cm}^{-3}]
      \label{eq:pop_part}
      \end{equation}
      \noindent
      where $K_2=K_1 \gamma_\mathrm{br}^{n_2-n_1}$ and $\gamma=E/m c^2$, where $m$ is the electron mass and $E$ its energy. These electrons radiate up to the X-ray range through the synchrotron process, and then re-interact with their own emitted photons by inverse Compton (IC) scattering, which is the so-called synchrotron self-Compton process. This synchrotron emission comes from a population of electrons different from those producing the radio-IR emission of the extended jet.
      
      The SSC model has 8 significant parameters. The macrophysics processes are described by the magnetic field $B$, the radius of the emitting blob $r_b$ and the Doppler factor $\delta_b=\left[ \Gamma_b (1 - \beta_b \cos{\theta}) \right]^{-1}$, where $\beta_b$ is the speed of the moving blob in $c$ unit, $\Gamma_b=(1-\beta_b^2)^{-1/2}$ is the blob Lorentz factor and $\theta$ is the viewing angle. The radiative processes are parametrized by the description of the population of emitting particles, with the parameters $K_1$, $\gamma_\mathrm{br}$, $\gamma_c$, $n_1$ and $n_2$ from Eq.~\eqref{eq:pop_part}. The value of $\gamma_\mathrm{min}$ is not crucial for the interpretation of the spectral energy distribution (SED), nor is $\gamma_c$, although it can become very relevant in cases where the X-rays have a hard slope with a spectral differential index $\alpha < 1$ (in the common $f_\nu \propto \nu^{-\alpha}$ notation). All these parameters can be constrained when detailed spectral data are available for a wide frequency range.

      In the present case, the spectral coverage of the nucleus of M\,87 is sparse and we need to find other ways to constrain the parameters. One important constraint comes from the variability:

      \begin{equation}
      r_b < \frac{c \delta_b}{1+z} \Delta t_\mathrm{obs}
      \label{eq:rb_constraint}
      \end{equation}
      where $\Delta t_\mathrm{obs}$ is the variability timescale in the observer frame, implying $r_b/\delta_b \la 5 \times 10^{15}$\,cm for M\,87.

      The region of emission is then assumed to be close to the SMBH, to fulfill the variability constraint within magnetohydrodynamic (MHD) jet models. For instance, \citet{2006MNRAS.368.1561M} models the jet formation zone using general relativistic magnetohydrodynamic simulations, applicable to GRBs, AGNs as M\,87 and black hole X-ray binaries. He describes the broadening zone of the jet in the vicinity of the central black hole, and finds the Alfv{\'e}n surface at $\sim$50\,$r_g$. We assume that the emission zone is located slightly above this surface to allow shocks and Fermi acceleration processes to develop in the jet. The results of \citet{2006MNRAS.368.1561M} further constrain some of our parameters for a distance of $\sim$100\,$r_g$ from the SMBH, such as the value of the Lorentz factor $\Gamma_b \la 10$ of the plasma blobs, the magnetic field $B$, and the half-opening angle $\varphi(r)$ of the jet given by his Eq.~(24).
      
      The case of M\,87 is of particular interest since its jet is exceptionally well mapped in radio VLBI. \citet{2002NewAR..46..239B} observed the core of M\,87 in VLBI in February 1995 and March 1999, and showed that the opening angle increases quickly with decreasing distance to the core region, at the 0.01\,pc ($\sim$70\,$r_g$) scale. Such a widening at the base of the jet was also observed in Cen\,A at the 0.1\,pc ($\sim$19\,000\,$r_g$) scale by \citet{2006PASJ...58..211H} using the VLBI Space Observatory Programme. A broadening zone in the jet formation region is found as well in MHD simulations. Moreover, the recently detected short term TeV variability seems to exclude the extended and outer regions as the source for VHE emission in M\,87. This is also argued by \citet{2005ApJ...634L..33G} who present a modified leptonic model applied to M\,87 which takes into account a deceleration of the inner flow along the base of the jet. So it appears quite natural to assume that the VHE $\gamma$-rays are emitted in the core widened jet region.

      We can then imagine that there are blobs of plasma, harboring very high energy electrons and propagating in the widened jet formation zone, that are dragged along with the bulk jet outflow. In the case of misaligned objects such as M\,87, this can easily result in one blob moving along the line of sight and thus having about the same Lorentz factor as for blazars, allowing to reproduce the TeV emission in the framework of classic SSC models. However a model with a single relativistic blob moving and emitting exactly towards the observer would be statistically unlikely.

      \section{Multi-blob model
	\label{sec:multi_blobs}}
      
      A way to deal with this statistical issue is to assume that the emission zone is a spherical cap centered on the SMBH, limited by the sheath of the jet and filled with several similar homogeneous blobs. Consequently, as mentioned in \citet{2006Sci...314.1424A}, we can consider differential Doppler boosting in the jet formation zone, near the core region.

      This cap is located at a given distance $R_\mathrm{cap}$ from the SMBH, which is a new free parameter in our model. However, $R_\mathrm{cap}$ can be constrained by MHD simulations \citep[e.g.][]{2006MNRAS.368.1561M} if we assume that it is located slightly above the Alfv{\'e}n surface, which is at about 50--100\,$r_g$  from the SMBH. This surface is continuous, but does not need to be homogeneous. We model it with a pattern of several blobs, whose individual radii are typically smaller than in the case of the ``blob-in-jet'' scenario.
   
      \begin{figure} 
	\psfrag{theta}{\Huge $\theta$}
	\psfrag{phi}{\Huge $\varphi(r)$}
	\psfrag{dalpha}{\Huge $d\alpha$}
	\psfrag{vj}{\Huge $v_j$}
	\psfrag{0, 1, 4}{\Huge 0, 1, 4}
	\psfrag{2, 3}{\Huge 2, 3}
	\psfrag{5, 6}{\Huge 5, 6}
	\psfrag{v014}{\Huge $v_{0,1,4}$}
	\psfrag{v23}{\Huge $v_{2,3}$}
	\psfrag{v56}{\Huge $v_{5,6}$}
	\psfrag{zone delta_b > 1}{\Huge zone $\delta_b > 1$}
	\psfrag{jet axis}{\Huge jet axis}
	\psfrag{Rcap}{\Huge $R_\mathrm{cap}$}
	\psfrag{rb}{\Huge $r_b$}
	\resizebox{\hsize}{!}{\includegraphics{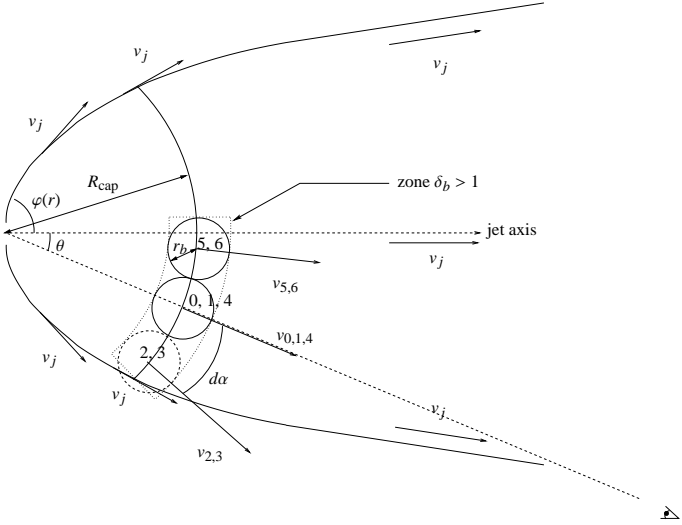}}
	\caption{Geometric side view of the jet formation zone (``on-blob'' case). $v_j$ represents the velocity of the jet, the blobs are numbered from 0 to 6, $v_{0 \rightarrow 6}$ are the velocities of the blobs, $\theta$ is the viewing angle with respect to the jet axis, $\varphi(r)$ is the opening angle, $r_b$ is the radius of an individual blob and $R_\mathrm{cap}$ is the distance of the blobs from the SMBH. In this sketch, the contributions to the total flux of the blobs n{\degr}2 and 3 would be rejected, since they lie outside the jet.
	}
	\label{fig:geom_multi_blobs}
      \end{figure}
      %
      %

      For the sake of simplicity, we assume in this zone the presence of 7 blobs, one central blob and 6 further blobs distributed on a hexagon, located at 100\,$r_g$ from the SMBH, with macroscopic parameters derived from \citet{2006MNRAS.368.1561M} as specified in Sect.~\ref{sec:bij}. This choice for the number of blobs is justified by the fact that the resulting diameter of the cap is of the same order of magnitude as the characteristic size of the emitting zone from previous studies.

      The smaller the radius of the individual blob, the more our model resembles a continuous zone model. The choice of discrete adjacent blobs leads to two extreme situations:
      
      \begin{itemize}
      \item ``Inter-blob'' case: the line of sight passes exactly through the gap between three blobs.
      \item ``On-blob'' case: the line of sight is exactly aligned with the velocity vector of the central blob, called n{\degr}0.
      \end{itemize}
      
      The Doppler factor for each blob n{\degr}$i$ (where $i \in \llbracket 0;6 \rrbracket $) is now defined as:
      
      \[
      \delta_b^i = \frac{1}{\Gamma_b (1 - \beta_b \cos \alpha_i)}
      \]
      where $\alpha_i$ is the angle between the velocity vector of blob n{\degr}$i$ and the line of sight. If the line of sight is between three blobs (``inter-blob'' case), then these blobs have the same Doppler factor and their contribution to the total flux is equal, while the 4 other blobs have contributions to the total flux that decrease with increasing blob radius $r_b$. If the line of sight is aligned with the velocity vector of the central blob (``on-blob'' case), then the highest Doppler factor is $\delta_b^0$. In that case the six other blobs have all the same Doppler factor, and although smaller than $\delta_b^0$ their individual contributions are not negligible in the total observed flux, especially if the seven blobs are all moving along in the same direction\footnote{which is the case for $r_b \leqslant r_g$.}.
      It should be noted that some models involve acceleration as the jet is collimated \citep[e.g.][]{2002ApJ...567..811M,2004ApJ...605..656V}, that is at the parsec scale. In such models, a gradient in flow velocity across the width of the jet can also be present, but is usually small compared to the radial velocity profile. Therefore we choose here to neglect this transverse gradient, which is of second order for our purpose, and we assume that all the blobs have the same Lorentz factor, although they are ejected at slightly different angles.
      
      Figure~\ref{fig:geom_multi_blobs} describes the geometry of our model in the ``on-blob'' case. The central blob is moving along the line of sight, and the six other blobs are each moving along a direction slightly different from the blob n{\degr}0 by an angle $d \alpha$. The angle $d \alpha$ is given by $d \alpha = 2 \arcsin\left(r_b/R_\mathrm{cap}\right)$. The viewing angle $\theta$ is defined as the angle between the line of sight and the jet axis for this multi-blob model. The individual radius $r_b$ assumed equal for all blobs is a free parameter of our model. Depending on the observed angle $\theta$, it can happen that in the simulation a blob moves outside the jet and is therefore neglected.

      \begin{figure} 
	\resizebox{\hsize}{!}{\includegraphics{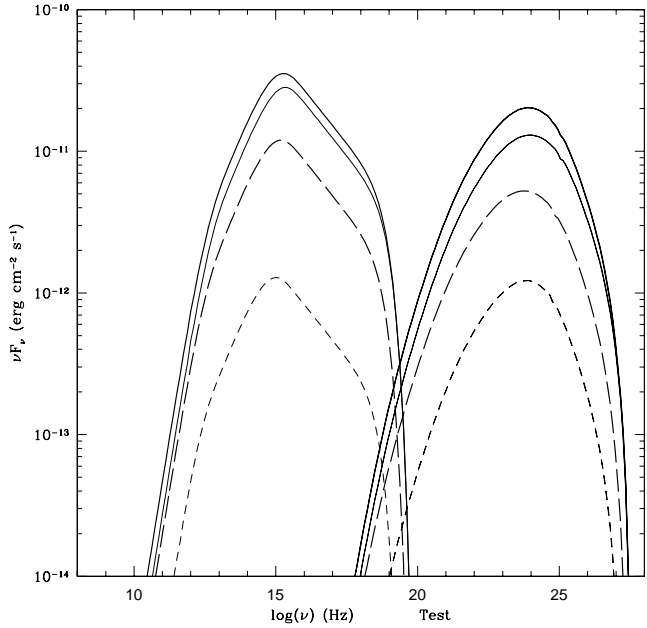}}
	\caption{Illustrative example of the SED of the multi-blob model. The thin solid line shows the contribution of the central blob, the most strongly beamed blob. The thin dashed line shows the contributions of each of the six other blobs, and the bold solid line shows the sum of the contributions of all the blobs in the ``on-blob'' case. The bold dashed line shows the sum of all the contributions in the ``inter-blob'' case. The closest blob to the line of sight ({\it thin solid line}) is overwhelmingly dominant for sufficiently large values of $r_b$.
	}
	\label{fig:SED_model_blobs_contrib}
      \end{figure}
      %
      %

      We can then compute the radiative transfer of each blob in its own source frame, as explained in \citet{2001A&A...367..809K}. For each blob, the seed photons for the inverse Compton scattering are those generated within the blob from synchrotron radiation. The total flux in the observer frame is the sum of the contribution of each blob. We neglect the contribution of the eventual blobs for which $\delta_b < 1$, which is possible if $\alpha_i > \arccos [(\Gamma_b-1)/(\Gamma_b \beta_b)]$.

      Figure~\ref{fig:SED_model_blobs_contrib} shows the SED of the contribution of the single central blob in the ``on-blob'' case ({\it thin solid line}), while the thin dashed line shows the contribution of the six adjacent blobs. Adjacent blobs have the same Doppler factor, so they all contribute equally to the SED. The bold solid line shows the total flux of the system, which is the sum of the contributions of the seven blobs. The bold dashed line shows the sum of all the contributions in the ``inter-blob'' case, where the line of sight is exactly between three blobs. If $r_b$ is sufficiently large, the closest blob to the line of sight in the ``on-blob'' case completely dominates the apparent flux. This can be the case in almost all the situations we study here.

      \section{Application to M\,87
	\label{sec:M87}}
      
      \subsection{The observed SED}
      
      To construct the SED of the core jet, we carefully selected the following data from the literature. We use $\gamma$-ray observations of 2004 and 2005 by H.E.S.S. \citep{2006Sci...314.1424A}, shown in black and gray points respectively in the next figures. The Whipple upper limit at 400\,GeV observed between 2000 and 2003 is taken from \citet{2004ApJ...610..156L}. The HEGRA point at 730\,GeV obtained in 1998 and 1999 is taken from \citet{2004NewAR..48..407B}.

      The {\it Chandra} data from the nucleus region obtained on April 20 and July 30, 2000 are taken from \citet{2005ApJ...627..140P} who conclude that the nuclear X-ray emission originates from the jet and could extract a 1 arcsecond nucleus spectrum thus excluding {\it HST}-1. The {\it XMM-Newton} data taken on June 19, 2000 were found in \citet{2001A&A...365L.181B}. It should be noted that the {\it XMM-Newton} data have the same spectral slope, but a higher flux density than the {\it Chandra} data, and they do include {\it HST}-1 within a $4\arcsec$ extraction region. \citet{2001ApJ...561L..51P} provide data taken by the Gemini North telescope at 10 microns in May 2001. The observations by the {\it HST} in the optical and UV bands in 1991 are reported in \citet{1996ApJ...473..254S}. The observations of the core of M\,87 by the VLA in the radio band and by the Palomar observatory in the optical band between 1979 and 1985 were found in \citet{1991AJ....101.1632B}.
      
      We also take into account upper limits in $\gamma$-ray by EGRET between 1991 and 1993 \citep{1994ApJ...426..105S}, and in UV by {\it EUVE} \citep{2000ApJ...535..615B}. All the other data are taken from the NASA/IPAC Extragalactic Database (NED).

      \begin{figure} 
	\resizebox{\hsize}{!}{\includegraphics{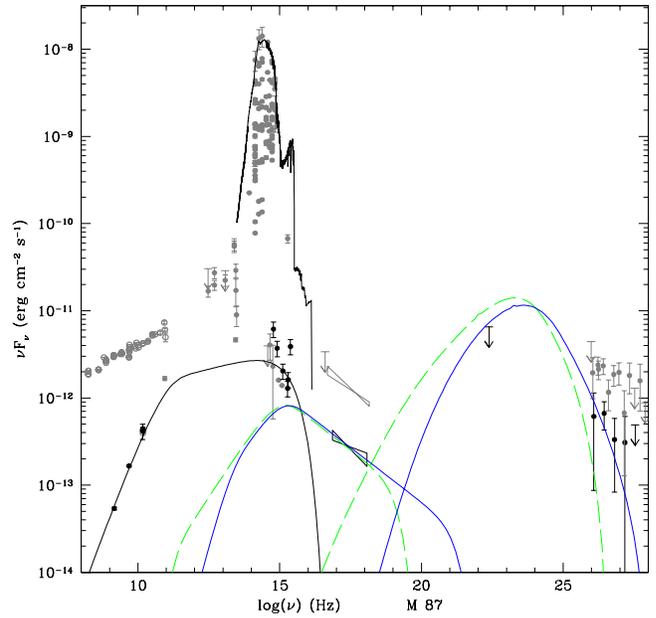}}
	\caption{Tentative modeling of the SED of M\,87 within a standard blob-in-jet scenario, with $\delta_b=3.07$ presented by the dashed green line. The solid blue line shows the SED of M\,87 emitted by a single blob moving along the line of sight in the jet formation region, with $\delta_b=8$, which can describe the data. The data points that bring direct constraints to our model are shown in black, the other less constraining points are represented in gray. The black line in the optical band shows the host galaxy, assumed to be elliptic. This contribution was computed using results from the code PEGASE \citep[][]{1997A&A...326..950F}. The radio data shown by gray empty circles, obtained from the NED and with fluxes $\sim$10$^{-12}$\,erg\,cm$^{-2}$\,s$^{-1}$, come from the extended kiloparsec-scale jet and radio lobes. The black line from radio to UV/X represents a model of the extended inner jet \citep[see][]{2001A&A...367..809K,2003A&A...410..101K}, with the corresponding radio data with fluxes $\sim$10$^{-13}$\,erg\,cm$^{-2}$\,s$^{-1}$ reported in black. The bump peaking at $\sim$10$^{15}$\,Hz is due to synchrotron emission and the bump peaking at $\sim$10$^{23}$\,Hz is due to inverse Compton, both from the VHE zone (see columns 1 and 2 of Table~\ref{tab:param} for the corresponding parameters). [{\it See the electronic edition of the Journal for a color version of this figure.}]
	}
	\label{fig:M87_SED_bij_broad}
      \end{figure}
      %
      %

      Although we do not have simultaneous data, the {\it Chandra} data of 2000 and the H.E.S.S. $\gamma$-ray data of 2004 both correspond to low states of activity in their own spectral range. We therefore associate them, assuming that they are representative of a typical low state. Indeed, regarding the X-ray data, the {\it Chandra} observations of 2000 correspond also to the lower state of activity recently published \citep{2003ApJ...599L..65P}. For the models, we choose to take into account the mean spectral slope of the {\it Chandra} data. The radio to optical/UV data are also not simultaneous with the $\gamma$-ray data, but this is not problematic since the radio contribution is thought to come from the extended jet, with characteristics different from the VHE emitting zone.

      \subsection{The SSC models}

      In the case of M\,87, the observation angle $\theta$ between the jet axis and the line of sight is at most $19\degr$ \citep{1999ApJ...520..621B}. The blob-in-jet model cannot describe correctly the VHE emission of the source, as it would require very high Doppler factor, which is not sustainable. This is illustrated in Fig.~\ref{fig:M87_SED_bij_broad}. The dashed green line presents the best solution for the SED of M\,87 within the blob-in-jet scenario described in Sect.~\ref{sec:bij} with $\delta_b=3.07$ assuming $\theta=19\degr$ (see column 1 in Table~\ref{tab:param} for the corresponding parameters). Greater values of $\delta_b$ are not allowed here because of the large value of $\theta$. We can obviously see that this model cannot account for the VHE emission. The well defined X-ray slope deduced from the observations strongly constrains the second index $n_2$ in the electron energy distribution (see Eq.~\eqref{eq:pop_part}) and significantly reduces the parameter space.

      \begin{figure} 
	\resizebox{\hsize}{!}{\includegraphics[angle=-90]{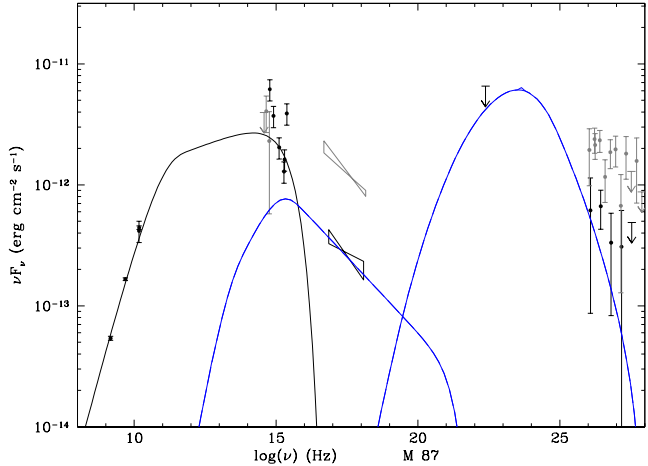}}
	\caption{SED of M\,87 within the multi-blob scenario for small blob radius ($r_b=1.5 \times 10^{13}$\,cm, see column 3 in Table~\ref{tab:param}). The two extreme ``on-blob'' and ``inter-blob'' cases are identical here because of the small value of $r_b$. [{\it See the electronic edition of the Journal for a color version of this figure.}]
	}
	\label{fig:M87_SED1}
      \end{figure}
      %
      %

      The solid blue line in Fig.~\ref{fig:M87_SED_bij_broad} presents the resulting SED emitted when one consider a single blob moving along the line of sight in the jet formation zone with $\delta_b = 8$. The corresponding parameters can be found in column 2 of Table~\ref{tab:param}, where $\theta$ is defined as the angle between the line of sight and the velocity vector of the single blob. Obviously this model describes the observations much better. However, as pointed out in Sect.~\ref{sec:bij}, it is based on an {\it ad hoc} assumption. Moreover, it is difficult to ``keep'' the generated IC bump below the EGRET upper limit although we assume a low state for the activity of the AGN.

      Throughout this paper, our results are not fits to the data, but rather solutions of models which are meant to describe best the data. Our purpose is to figure out whether our model can describe correctly the current available data for different objects. We do not intend to fine-tune the parameters of our model but to sort orders of magnitude out for these parameters.

      One SED of M\,87 generated within the multi-blob model is presented in Fig.~\ref{fig:M87_SED1}, with parameters very similar to the single blob model of Fig.~\ref{fig:M87_SED_bij_broad} (see column 3 in Table~\ref{tab:param}). Since the former is a generalization of the latter, the resulting spectrum is rather similar, as one would expect. In this case, the value of the individual blob radius $r_b$ is so small that all the blobs are moving close to the line of sight. The ``on-blob'' and the ``inter-blob'' cases give the same contribution to the SED and are overlaid in Fig.~\ref{fig:M87_SED1}.

      \begin{figure} 
	\resizebox{\hsize}{!}{\includegraphics[angle=-90]{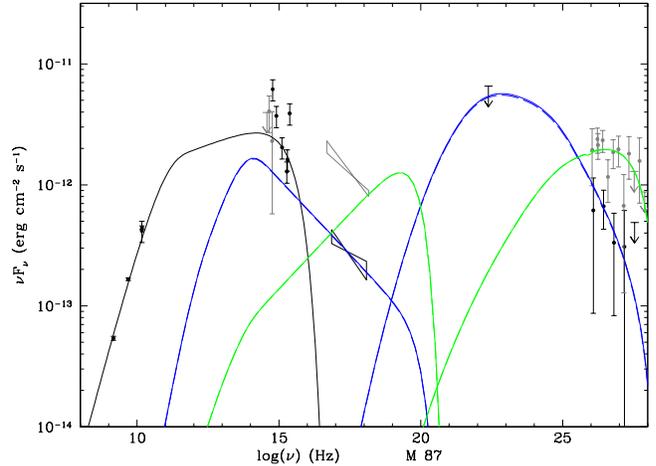}}
	\caption{SED of M\,87 within the multi-blob scenario, assuming a low magnetic field ($B=10$\,mG, see column 4 in Table~\ref{tab:param}) in blue lines. The solid lines show the ``on-blob'' case while the dashed lines represent the ``inter-blob'' case. The green lines show a solution for the high state observed by H.E.S.S. in 2005, with the same parameters as for the observations of 2004, except for $r_b = 8.0 \times 10^{13}$\,cm, $K_1 = 2.2 \times 10^4$\,cm$^{-3}$ and $n_2 = 2.5$. [{\it See the electronic edition of the Journal for a color version of this figure.}]
	}
	\label{fig:M87_SED2}
      \end{figure}
      %
      %

      The blob radius is rather small in this case, resulting in a VHE emitting zone smaller than the Schwarzschild radius. It should be noted that features small compared to the Schwarzschild radius, possibly responsible for the VHE emission, can develop beyond the Alfv{\'e}n surface due to turbulence or reconfined shocks, but this issue is beyond the scope of this work.

      However it is commonly believed that the size of the VHE emitting zone cannot be much smaller than the Schwarzschild radius which is a natural scale for the processes in the vicinity of the SMBH. Moreover the emitting zone must be large enough to allow the acceleration of particles to develop. Very small blobs may disappear rapidly, in $\sim 10$\,minutes due to adiabatic expansion which is especially important in the broadened zone of the jet. However a long, stable emission is possible, even from small blobs. The emitting zone can be located at a stable stationary shock front, above the Alfv\'en surface. It initiates the acceleration and thus the radiation of particles of a large number of small blobs continuously crossing the shock, thus providing a quiescent background of VHE emission. Density fluctuations in the injection of material could then generate flares as seen at VHE. In fact, the only problem with small blobs is that in this case the paving of the jet is not complete because of the discretization applied in our code.

      In order to be more conservative and to fulfill the constraint $r_b \ga r_g$, we analyze another possibility with a low magnetic field. It is presented in Fig.~\ref{fig:M87_SED2} in blue lines with associated parameters in column 4 of Table~\ref{tab:param}. In that case, the result predicted by MHD models with a strong magnetic field in the vicinity of the central engine is not strictly fulfilled. However a local decrease of the magnetic field can be achieved by a simple expansion of the emitting zone. This solution may appear preferable for conservative reasons on the size of the emitting zone.

      A satisfactory solution for the high state observed by H.E.S.S. in 2005 is also possible in this case within the multi-blob scenario as shown in Fig.~\ref{fig:M87_SED2} ({\it green lines}). Interestingly we predict a radical change of the X-ray regime. Clearly the parameters are not well constrained here due to the lack of any simultaneous data, especially in the X-rays. However this illustrates the capability of the multi-blob model to generate spectra that are sufficiently hard in the TeV range to reproduce the most recent H.E.S.S. data. A slight modification of standard SSC models for TeV blazars appears therefore successful and can account for all VHE data on the radiogalaxy M\,87 that are available up to now.

      \begin{table*} 
	\caption{Parameters used in the different generated SEDs.}
	\label{tab:param}
	\centering
	\begin{tabular}{l|c|c|cccccc}
	  \hline \hline
	  Model & Blob-in-jet & Single blob & \multicolumn{6}{c}{Multi-blob}\\
	  &       & in broadened zone & & & & & & \\ 
	  Object & M\,87 & M\,87 & M\,87 & M\,87 & 3C\,273 & Cen\,A & Cen\,A & PKS\,0521$-$36\\
	  Figure & \ref{fig:M87_SED_bij_broad} ({\it green}) & \ref{fig:M87_SED_bij_broad} ({\it blue}) & \ref{fig:M87_SED1} & \ref{fig:M87_SED2} & \ref{fig:3C273_SED} & \ref{fig:CenA_SED} ({\it blue}) & \ref{fig:CenA_SED} ({\it green}) & \ref{fig:pks0521_SED}\\
	  \hline
	  $\delta_b$ & 3.07 & 8 & -- & -- & --& -- & -- & --\\
	  $\Gamma_b$ & -- & -- & 4.1 & 10.0 & 7.4 & 8.14 & 20.0 & 1.5\\
	  $\theta$ & $19\degr$ & $1\degr$ & $15\degr$ & $15\degr$ & $15\degr$ & $25\degr$ & $25\degr$ & $25\degr$\\
	  $R_\mathrm{cap}$ [$r_g$] & -- & -- & 100.0 & 100.0 & 100.0 & 100.0 & 100.0 & 100.0\\
	  $B$ [G] & 1.0 & 0.5 & 0.5 & 0.01 & 3.0 & 2.0 & 10.0 & 1.0\\
	  $r_b$ [cm] & $1.2 \times 10^{14}$ & $4.0 \times 10^{13}$ & $1.5 \times 10^{13}$ & $2.8 \times 10^{14}$ & $2.0 \times 10^{15}$ & $1.0 \times 10^{14}$ & $8.0 \times 10^{13}$ & $9.0 \times 10^{14}$\\
	  $K_1$ [cm$^{-3}$] & $1.5 \times 10^7$ & $3.5 \times 10^7$ & $7.7 \times 10^7$ & $1.8 \times 10^4$ & $1.8 \times 10^6$ & $9.0 \times 10^7$ & $4.0 \times 10^4$ & $3.0 \times 10^6$\\
	  $n_1$ & 2.0 & 2.0 & 2.0 & 1.5 & 2.0 & 2.0 & 2.0 & 2.0\\
	  $n_2$ & 3.5 & 3.5 & 3.5 & 3.5 & 4.1 & 3.0 & 3.5 & 2.5\\
	  $\gamma_\mathrm{min}$ & $10^2$ & $10^3$ & $10^3$ & $10^3$ & $1$ & $3.0 \times 10^2$ & $10^3$ & $10^3$\\
	  $\gamma_\mathrm{br}$ & $10^4$ & $10^4$ & $10^4$ & $10^4$ & $1.6 \times 10^3$ & $5.0 \times 10^2$ & $3.5 \times 10^5$ & $5.0 \times 10^4$\\
	  $\gamma_c$ & $10^6$ & $10^7$ & $10^7$ & $10^7$ & $10^6$ & $4.0 \times 10^3$ & $6.0 \times 10^6$ & $10^6$\\
	  \hline
	\end{tabular}
      \end{table*}
      %
      %

      \section{Predictions for other radiogalaxies with optical/X-ray extended jets
	\label{sec:predictions}}

      We now apply the SSC multi-blob model to different sources that have the peculiarity to show an extended optical or X-ray jet, which suggests at least a moderate beaming towards the observer as in the case of M\,87. This allows to confront the multi-blob scenario to other types of AGNs and to predict whether these sources can be detectable at VHE or not by present \v{C}erenkov arrays. We choose three AGNs not belonging to the genuine blazar class, presented here with increasing viewing angles, such that their fluxes are less and less boosted by relativistic effects. We stress that not all the data presented here are simultaneous and that the sources undergo large variations.

      \subsection{3C\,273}

      3C\,273 \citep[z=0.158,][]{1992ApJS...83...29S} is the first quasar that was identified as a high-redshift object \citep[][]{1963Natur.197.1040S} and the best studied. It hosts a SMBH whose mass is at least $\sim$2.0$ \times 10^9 M_{\sun}$ as inferred from studies of Balmer lines \citep[][]{2005A&A...435..811P} and its maximum acceptable viewing angle is about 15\degr \citep[][]{1985ApJ...289..109U}.

      Blazars display featureless X-ray contribution but radiogalaxies can have a much more complex environment at low energies \citep[e.g.][]{2006ApJ...642..113G}. However since the purpose here is to model the non-thermal contributions of these objects, we decided to consider the X-ray contribution as dominated by the jet emission, keeping only a feature in soft X-ray as a signature of the accretion disk \citep{2007ApJ...659..235G}.

\begin{figure} 
	\resizebox{\hsize}{!}{\includegraphics[angle=-90]{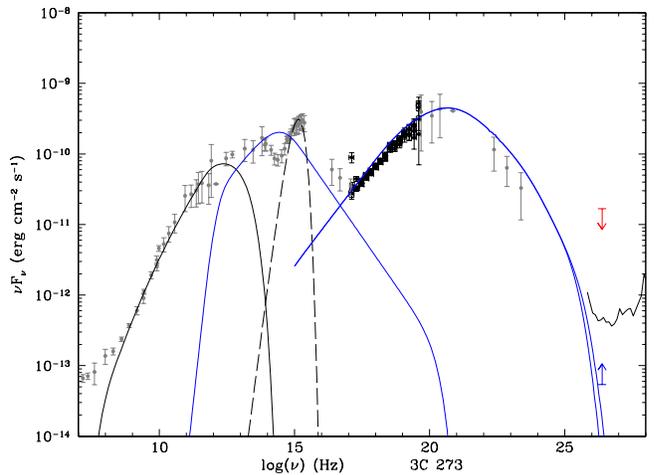}}
	\caption{SED of 3C\,273 with anticipated VHE flux. The solid blue line shows the average of the ``on-blob'' and ``inter-blob'' cases (see column 5 in Table~\ref{tab:param} for the corresponding parameters). The dashed line represents a simple blackbody model to illustrate the contribution of the big blue bump component. The upper limit obtained in 2005 with the H.E.S.S. telescope array is shown in red. The blue lower limit shows the expected CTA sensitivity in 50\,h of observation. [{\it See the electronic edition of the Journal for a color version of this figure.}]
	}
	\label{fig:3C273_SED}
      \end{figure}

      Since this source is highly variable, one needs to be careful to select simultaneous data. The X-ray data presented here are {\it Beppo}SAX observations\footnote{see \url{http://www.asdc.asi.it/blazars/}.} taken from \citet{2002babs.conf...63G}. We report in red in Fig.~\ref{fig:3C273_SED} the upper limit at $3\sigma$ obtained by H.E.S.S. in 2005 \citep{2005A&A...441..465A}. All the other data points presented in gray are taken from \citet{1999A&AS..134...89T} who report an average spectrum compiled from 30 years of observations. Observations by {\it Beppo}SAX on January 13, 15, 17 and 22, 1997 lie within the observation period of {\it CGRO} taken by COMPTEL and EGRET between December 10, 1996 and January 28, 1997 \citep{2000A&A...354..513C} which are also reported in \citet{1999A&AS..134...89T}. We have thus simultaneous data for the X/$\gamma$-ray bump in one of the highest state, which puts an important constraint on the models.

      %
      %

      The nature of the X-ray emission of 3C\,273 is still an unresolved issue. The high frequency bump, thought to be probably due to IC emission, would present a peak at a rather low frequency compared to other AGNs. This implies that either 3C\,273 does not emit at detectable VHE levels, or that the nature of this bump is synchrotron, implying then the presence of a hypothetical IC bump at ultra high energies.

      Assuming that the hard X-ray emission is indeed due to inverse Compton emission, the solid blue line in Fig.~\ref{fig:3C273_SED} shows the SED of 3C\,273 with the predicted VHE flux (see column 5 of Table~\ref{tab:param} for the corresponding parameters). Modeling the SED of 3C\,273, we note that changing the value of $\gamma_\mathrm{min}$ dramatically affects the X-ray flux by increasing it with decreasing $\gamma_\mathrm{min}$, and hence could explain some X-ray flares, as also suggested by \citet{2006ApJ...653L...5G} in the case of external inverse Compton emission on the Cosmic Microwave Background. The value of $\gamma_\mathrm{min}$ is strongly constrained here by the precise shape of the X-ray spectrum. In the following figures, the V-shaped curve at VHE ($\sim$10$^{26}$\,Hz) shows the H.E.S.S. sensitivity limit for a detection of $5\sigma$ in 50\,h of observation for a source at a mean zenith angle of 30\degr. The expected sensitivity of the next generation CTA project of $\sim$0.1\% Crab flux at 1\,TeV for 50\,h of observation is shown as a blue lower limit.

      Since the results for the ``on-blob'' and ``inter-blob'' scenarii are not very different in this case, we show with a blue line the average of the two. We should stress that low frequency data are not simultaneous with the X/$\gamma$-ray data that we selected. The modeled synchrotron bump has a higher flux density than the optical data because we are dealing with a high state of activity in $\gamma$-rays as reported by \citet{2000A&A...354..513C}. We only predict a very marginal detection of 3C\,273 by H.E.S.S. in its low energy range, depending on the energy threshold \citep[but see also][]{2006ApJ...653L...5G}. Furthermore it should be recalled that \citet{2000A&A...354..513C} report an active state in $\gamma$-rays and \citet{2006A&A...451L...1T} a high level of the non-thermal emission at the epoch of the data we are considering. So even in a high state we do not expect a strong level of VHE $\gamma$-ray within our scenario.

      A strong detection of 3C\,273 at VHE with the current generation of \v{C}erenkov arrays would be difficult to explain within our SSC nuclear scenario. A possibility would be to invoke disparity among the various emitting plasma blobs. Our model shows the presence of a well peaked inverse Compton bump; different magnetic fields or electron energy distributions among the blobs could result in a tail of the IC bump at VHE that could account for a VHE detection. An alternative would be an extended emission due to external inverse Compton radiation, which is then expected to be not very variable. In all cases, further observations with {\it GLAST} (10\,keV--300\,GeV) and H.E.S.S.\,II, which will extend the spectral domain of H.E.S.S.\,I down to $\sim$20\,GeV with a better sensitivity, are required to disentangle the different plausible scenarii.

      \subsection{Cen\,A}
      
      Cen\,A is the nearest radiogalaxy \citep[z=0.0018,][]{1978PASP...90..237G} and one of the best studied. The presence of an AGN in Cen\,A is evident from the radio-band to the $\gamma$-rays. Observations from {\it CGRO} \citep[][]{1995ApJ...449..105K} show a bump that seems to peak at $\sim$200\,keV as pointed out by \citet[][]{1998A&A...330...97S}. We thus have a real constraint for the parameters of our model, particularly with regard to the description of the population of electrons. As in the case of 3C\,273, the IC bump peaks at a rather low energy, leading \citet{2001MNRAS.324L..33C} to note that Cen\,A could be a misaligned low-energy peaked BL\,Lac (LBL) object. The value of the viewing angle of the jet is still a controversial issue. For instance, \citet{1998AJ....115..960T} claim $\theta \sim 50\degr$--$80\degr$ for the parsec-scale jet, whereas \citet{2003ApJ...593..169H} find $\theta \sim 15\degr$ for the 100\,pc scale jet. We choose here to take an intermediate value of $\theta \sim 25\degr$ (See \citet{2006PASJ...58..211H} for a discussion about the different values for the viewing angle of Cen\,A found in the literature). The SMBH mass inferred from gas kinematical analysis using a [S\,III] line is $\sim$1.1$ \times 10^8 M_{\sun}$ \citep[][]{2006A&A...448..921M}, but see also \citet{2006ApJ...643..226H} who give $M_\mathrm{BH} \sim 6 \times 10^7 M_{\sun}$ using a [Fe\,II] line.

      \begin{figure} 
	\resizebox{\hsize}{!}{\includegraphics[angle=-90]{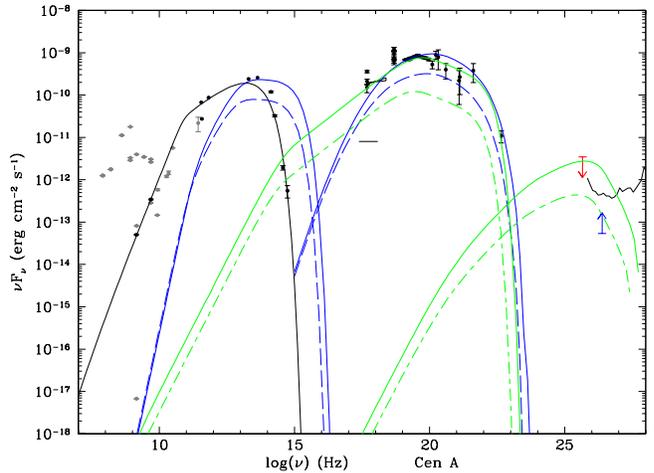}}
	\caption{SED of Cen\,A within the multi-blob scenario. Solid lines show the ``on-blob'' emission; the ``inter-blob'' cases are represented in dashed lines. In blue we show a model assuming that the soft $\gamma$-ray data are inverse Compton emission, while in green we assume a synchrotron emission to describe them (see columns 6 and 7 respectively in Table~\ref{tab:param} for the corresponding parameters). [{\it See the electronic edition of the Journal for a color version of this figure.}]
	}
	\label{fig:CenA_SED}
      \end{figure}
      %
      %

      The data sample chosen here is almost the same as in \citet{2001MNRAS.324L..33C}. The data were retrieved carefully to take only the nucleus into account. The strongly constraining {\it CGRO}/COMPTEL $\gamma$-ray data are taken from \citet{1998A&A...330...97S}; \citet{2006ApJ...641..801R} provided {\it RXTE} and {\it INTEGRAL} data; {\it HST}/NICMOS and WFPC2 data, which we have carefully dereddened, are from \citet{2000ApJ...528..276M}\footnote{The reddening correction was applied using the factors carried out by \citet{2000ApJ...528..276M} themselves.}; SCUBA data at 800\,$\mu$m are taken from \citet{1993MNRAS.260..844H}; {\it ISO} and SCUBA (450\,$\mu$m and 850\,$\mu$m) data are from \citet{1999A&A...341..667M}; VLA data are from \citet{1983ApJ...273..128B}. \citet{2004ApJ...612..786E} report X-ray observations by {\it XMM-Newton} on February 2 and 6, 2001 and by {\it Chandra} on May 9 and 21, 2001 with a photon index $\Gamma=2$ for the parsec-scale jet component. Data from the NED are shown as non-constraining points ({\it in gray}) for comparison. The H.E.S.S. upper limit based on observations in 2004 with a live time exposure of 4.2\,h is reported in red \citep{2005A&A...441..465A}.

      We should also point out that as Cen\,A harbors a strongly absorbing dust lane, and because this source is extremely close and well resolved, the X-ray data should be then taken only as upper limits. The contribution of the nuclear jet might be contaminated by the accretion disk of the AGN and by the X-ray binaries that are resolved in this object. In this case we would have only poor constraints on the emission process. We assume here that all our selected high energy data come from SSC processes.

      Figure~\ref{fig:CenA_SED} shows the SED of Cen\,A applying the multi-blob model in two cases, ({\it i}) assuming that the $\gamma$-ray peak observed by {\it CGRO} is inverse Compton radiation ({\it solid and dashed blue lines}) or ({\it ii}) assuming it to be synchrotron ({\it solid and short dashed--long dashed green lines}).

We should also point out that the previous study by \citet{1998A&A...330...97S} reports a variability in soft $\gamma$-ray of about 10~days, implying $r_b < 2 \times 10^{17}$cm (see Eq.~\eqref{eq:rb_constraint}), which is well satisfied by our parameters.

      Given the results of our model in the first scenario with an IC bump in soft $\gamma$-rays ({\it blue lines}), the SSC emission of the central region would definitely not provide a flux sufficiently high to be detectable at VHE (see column 6 in Table~\ref{tab:param}), at least for a SSC emission dominated by the nucleus. This holds even in the case of huge variations of the nuclear emission. This conclusion concurs with \citet{2003ApJ...597..186S}, who do not expect SSC emission by the nucleus or by the base of the jet of Cen\,A, but do expect VHE emission that could be detectable by current imaging atmospheric \v{C}erenkov telescopes facilities in the case of an external inverse Compton emission process on the host galaxy photon field. For \citet{2006MNRAS.371.1705S}, many Fanaroff-Riley type I (FR\,I) radiogalaxies like Cen\,A could be TeV sources for which the weak nuclear $\gamma$-ray emission would be absorbed and re-processed by inverse Compton on the starlight photon field, thus generating an isotropic $\gamma$-ray halo. In our model, the lack of simultaneous data prevents us from further constraining the synchrotron bump, which has here a higher flux density than the selected data since we are considering a state of high $\gamma$-ray activity.

      In solid (``on-blob'' case) and short dashed--long dashed (``inter-blob'' case) green lines in Fig.~\ref{fig:CenA_SED}, we present a SED of Cen\,A assuming now that the soft $\gamma$-ray peak is of synchrotron origin. In this case (see column 7 in Table~\ref{tab:param}), we expect a detection of the core of Cen\,A at VHE by the H.E.S.S. telescope array within 50\,h of observation. It should be noted that \citet{2001ApJ...549L.173B} also predicted the synchrotron bump to be in the soft $\gamma$-ray range and the inverse Compton bump to peak around 1\,TeV in the context of SSC models, which comforts our latter model.

      \subsection{PKS\,0521$-$36}
      
      PKS\,0521$-$36 is a FSRQ object with an optical jet located at $z=0.055$ \citep[][]{1985AJ.....90.2207K}. The central SMBH has a mass of $\sim$3.3$ \times 10^8 M_{\sun}$ \citep[][]{2005ApJ...631..762W}. \citet{1996ApJ...459..169P}, and more recently \citet{2002AJ....124..652T}, mention the absence of superluminal motions in its jet, contrary to the case of 3C\,273, implying that the beaming effect is much less important and thus strengthening their result on the viewing angle. Indeed the only constraint on the jet orientation comes from \citet{1996ApJ...459..169P} who deduce $\theta \simeq 30\degr \pm 6\degr$ from SSC models. We should also note that PKS\,0521$-$36 seems to oscillate between a Seyfert-like and a BL\,Lac state \citep[e.g.][]{1981A&A...103L...1U}, making this source difficult to interpret within a pure non-thermal scenario, especially since we are confronted with non-simultaneous data.

\begin{figure} 
	\resizebox{\hsize}{!}{\includegraphics[angle=-90]{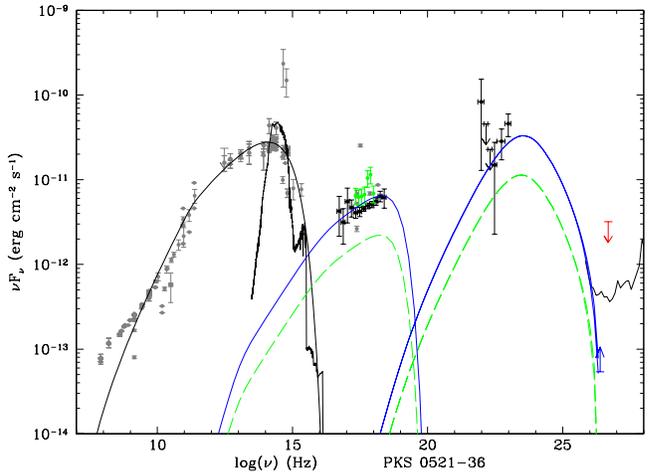}}
	\caption{SED of PKS\,0521$-$36 with anticipated VHE flux modeled in the multi-blob framework. The solid blue line is the ``on-blob'' case while the dashed green line shows the ``inter-blob'' emission (see column 8 in Table~\ref{tab:param} for the corresponding parameters). The CANGAROO upper limit is reported in red. [{\it See the electronic edition of the Journal for a color version of this figure.}]
	}
	\label{fig:pks0521_SED}
      \end{figure}

      {\it Beppo}SAX observed PKS\,0521$-$36 on October 10, 1998 \citep{2002babs.conf...63G} ({\it black points} in Fig.~\ref{fig:pks0521_SED}), and the {\it Swift}/XRT measurements ({\it green points}) taken on May 26, 2005 were obtained through the Online Analysis Tool\footnote{see \url{http://www.asdc.asi.it}}. The data points in gray are from the NED. We report in red the upper limit at $2\sigma$ obtained in 89\,h from observations by CANGAROO between 1993 and 1996 \citep{1998A&A...337...25R}. We further used the EGRET data between 30\,MeV and 500\,MeV from \citet{1999ApJS..123...79H} and taken between July 12, 1994 and August 01, 1994. The blazar PKS\,0521$-36$ is associated with the source 2EG\,J0524$-$3630 in the Second EGRET Catalog \citep{1995ApJS..101..259T}, but during cycle 4 this source was found to lie outside the 99\% confidence contour of EGRET. However, like \citet{2006A&A...453..829F}, we assume in this work the identification with PKS\,0521$-$36 to be valid, which is also pointed out by \citet{2002ApJ...579..136T}.

      %
      %

      Figure~\ref{fig:pks0521_SED} presents the SED of PKS\,0521$-$36 with the anticipated VHE emission (see column 8 in Table~\ref{tab:param} for the corresponding parameters). It seems unlikely that the X-rays are due to inverse Compton radiation since the inverse Compton scattering would be with photons from the radio/optical contribution, which is not very variable, coming from an extended part of the jet. Since the X-rays show a high degree of variability, they most certainly come from a compact region and are of synchrotron origin. If the X-ray emission is truly due to the synchrotron process, we predict that this BL\,Lac object should be marginally detectable by the present H.E.S.S. telescope array, and easily detectable by H.E.S.S.\,II and by the next generation of \v{C}erenkov arrays, such as the CTA project, which will detect sources down to $\sim$0.1\% of the Crab flux. Furthermore, if PKS\,0521$-$36 remains undetected at VHE, a misidentification with the EGRET source should be considered.

      \section{Discussion and implications on the AGN unification scheme
	\label{sec:unification}
      }

      When applied to AGNs belonging to very different classes, the multi-blob model deduces very similar properties for the size of the TeV emitting zones and the values of the magnetic field. Furthermore the inferred bulk Lorentz factors $\Gamma_b$ can usually remain below a value of 10, thus reconciling SSC models with (GR)MHD models, except for the interpretation of the $\gamma$-ray bump of Cen\,A in terms of synchrotron emission (see the green lines in Fig.~\ref{fig:CenA_SED}). A rather unified picture seems therefore to come out from the analysis.
      
      In our scenario, we locate the X- and $\gamma$-ray emitting region in the jet formation zone, with opening angle larger than in the global VLBI radio zone, which is more distant from the core and mainly located in a region where the jet is more collimated. This is somewhat reminiscent of a proposal made several years ago by \citet{1993ApJ...416..118C} for unifying X-ray and radio selected BL\,Lac objects. \citet{1993ApJ...416..118C} showed that such a picture can be coherent with statistics of the BL\,Lac population.

      Since our model accounts for the observation with rather low bulk Lorentz factors and with large effective opening angles, we have found a way out of the problem of statistics on the number of detected TeV sources invoked by \citet{2006ApJ...640..185H}, and we can reconcile beamed and unbeamed sources. One important consequence of our proposal is that bright radio BL\,Lacs should be TeV emitters. This can be tested by further observations.

      Indeed, \citet{2007Ap&SS.tmp..190P} shows that the common historical unified LBL/HBL scheme (the so-called ``blazar sequence'') seems to be ruled out by the discovery of ``outliers'' low-power LBL and high-power HBL sources. Hence it does not seem impossible that objects that are very different at first sight, like radiogalaxies such as Cen\,A, could be (faint) VHE emitters. Furthermore the fact that M\,87 has already been detected in the TeV range is encouraging for a future detection of such radiogalaxies. In the case of 3C\,273, this object would rather be a misaligned LBL-like object. However one should stress that the fact that mainly HBL have been detected up to now at VHE is certainly only a selection effect. Thus a TeV detection of 3C\,273 would not be very surprising.
      
      The scenario we propose here has an additional interest in the sense that it allows to solve the long standing paradox of the apparent absence of high superluminal motion at the base of radio jets of TeV BL\,Lacs. In our model, some X- and $\gamma$-ray emitting plasma blobs are moving close to the line of sight, thus allowing for instance Cen\,A and PKS\,0521$-$36 to be potentially seen at VHE, while standing in misaligned extended jets.

      This statement agrees well with the most recent studies by \citet{2006MNRAS.369.1287G,2007MNRAS.377..446G} who show that viewing angles, opening angles and Lorentz factors of (sub-)parsec scale jets evaluated by radio observations are usually underestimated, thus reconciling absent superluminal motions inferred from radio observations with high Lorentz factors required by different families of TeV emission models. \citet{2006astro.ph..3728W} also underlines the possibility to reconcile various contradictory observations by considering jets with opening angles of a few degrees. Early studies by \citet{1995ApJ...447..103D} on $\gamma$-ray observations of AGNs with EGRET also indicate that strong beaming is not required to account for TeV blazars observations.

      Two different variability time scales appear within our mutli-blob scenario. The short time scale is related to the characteristic size of individual blobs, as already discussed. A longer time scale occurs in the case of rotating jets with helicoidal magnetic field. This induces a rotation of the cap, and the lag between ``on-blob'' and ``inter-blob'' emission corresponds to a long term $\gamma$-ray variability. From MHD models, we infer a characteristic rotation time of the magnetic field of the order of one year in the observer frame for M\,87. This modulation could explain the variation of the VHE emission observed between low and high states in 2004 and 2005. This would imply some periodicity in the TeV emission of AGNs, but on timescales hitherto not explored.

      \section{Conclusion
	\label{sec:conclusion}
      }
      
      We have presented a SSC model to interpret VHE emission of M\,87 as well as other misaligned sources with extended optical/X-ray jet. This model accounts in a simple way for a differential Doppler boosting effect by modeling the emission of several blobs of plasma located in the broadened formation zone of the jet close to the SMBH, just beyond the Alfv{\'e}n surface predicted by MHD models.

      Our scenario provides a reasonable interpretation of the H.E.S.S. VHE observations of M\,87 and provides the possibility to extend standard leptonic models of TeV blazars to other types of AGNs. However, we do not exclude other leptonic or hadronic models. For instance, \citet{2007arXiv0704.3282N} recently interpreted the H.E.S.S. observations from 2005 of M\,87 by invoking acceleration and radiation of electrons in the black hole magnetosphere, which is another kind of leptonic model. Hadronic models also cannot be excluded as efficient particle acceleration processes can occur in the close surroundings of the black hole.

      More observations are needed to constrain the emission models and especially to distinguish between hadronic and leptonic scenarii. The upcoming {\it GLAST} mission and the H.E.S.S.\,II project will certainly help to understand the mechanisms at work in the AGNs by exploring spectral ranges below TeV, which is decisive to constrain the shape of the inverse Compton bump. Moreover, the lack of genuine simultaneous multiwavelength campaigns on M\,87 needs to be filled, especially since this source is known to be variable at small timescales in VHE.

      Several types of active nuclei are potential emitters of VHE photons with predicted TeV fluxes detectable by present \v{C}erenkov arrays like H.E.S.S. and MAGIC, or by the next generation of instruments such as the CTA project. Such data will be crucial to test AGN unifying schemes.

      \begin{acknowledgements}
	We are grateful to the anonymous referee for useful comments. J.-P.~L. would like to thank Dr.~A.~Djannati-Ata{\"i}, Dr.~S.~Pita and Dr.~A.~Zech for useful discussions.
	
	This research has made use of the NASA/IPAC Extragalactic Database (NED) which is operated by the Jet Propulsion Laboratory, California Institute of Technology, under contract with the National Aeronautics and Space Administration.
      \end{acknowledgements}

      \bibliographystyle{aa}  
      \bibliography{lenain_M87_paper}

\begin{thebibliography}{84}
\expandafter\ifx\csname natexlab\endcsname\relax\def\natexlab#1{#1}\fi

\bibitem[{{Aharonian} {et~al.}(2003){Aharonian}, {Akhperjanian}, {Beilicke},
  {Bernl{\"o}hr}, {B{\"o}rst}, {Bojahr}, {Bolz}, {Coarasa}, {Contreras},
  {Cortina}, {Denninghoff}, {Fonseca}, {Girma}, {G{\"o}tting}, {Heinzelmann},
  {Hermann}, {Heusler}, {Hofmann}, {Horns}, {Jung}, {Kankanyan}, {Kestel},
  {Kohnle}, {Konopelko}, {Kornmeyer}, {Kranich}, {Lampeitl}, {Lopez}, {Lorenz},
  {Lucarelli}, {Mang}, {Meyer}, {Mirzoyan}, {Moralejo}, {Ona-Wilhelmi},
  {Panter}, {Plyasheshnikov}, {P{\"u}hlhofer}, {de los Reyes}, {Rhode},
  {Ripken}, {Rowell}, {Sahakian}, {Samorski}, {Schilling}, {Siems},
  {Sobzynska}, {Stamm}, {Tluczykont}, {Vitale}, {V{\"o}lk}, {Wiedner}, \&
  {Wittek}}]{2003A&A...403L...1A}
{Aharonian}, F., {et~al.} (HEGRA Collab.) 2003, \aap, 403,
  L1

\bibitem[{{Aharonian} {et~al.}(2005){Aharonian}, {Akhperjanian}, {Bazer-Bachi},
  {Beilicke}, {Benbow}, {Berge}, {Bernl{\"o}hr}, {Boisson}, {Bolz}, {Borrel},
  {Braun}, {Breitling}, {Brown}, {Chadwick}, {Chounet}, {Cornils},
  {Costamante}, {Degrange}, {Dickinson}, {Djannati-Ata{\"i}}, {O'C.~Drury},
  {Dubus}, {Emmanoulopoulos}, {Espigat}, {Feinstein}, {Fontaine}, {Fuchs},
  {Funk}, {Gallant}, {Giebels}, {Gillessen}, {Glicenstein}, {Goret},
  {Hadjichristidis}, {Hauser}, {Heinzelmann}, {Henri}, {Hermann}, {Hinton},
  {Hofmann}, {Holleran}, {Horns}, {Jacholkowska}, {de Jager}, {Kh{\'e}lifi},
  {Komin}, {Konopelko}, {Latham}, {Le Gallou}, {Lemi{\`e}re},
  {Lemoine-Goumard}, {Leroy}, {Lohse}, {Martin}, {Martineau-Huynh},
  {Marcowith}, {Masterson}, {McComb}, {de Naurois}, {Nolan}, {Noutsos},
  {Orford}, {Osborne}, {Ouchrif}, {Panter}, {Pelletier}, {Pita},
  {P{\"u}hlhofer}, {Punch}, {Raubenheimer}, {Raue}, {Raux}, {Rayner}, {Reimer},
  {Reimer}, {Ripken}, {Rob}, {Rolland}, {Rowell}, {Sahakian}, {Saug{\'e}},
  {Schlenker}, {Schlickeiser}, {Schuster}, {Schwanke}, {Siewert}, {Sol},
  {Spangler}, {Steenkamp}, {Stegmann}, {Tavernet}, {Terrier}, {Th{\'e}oret},
  {Tluczykont}, {Vasileiadis}, {Venter}, {Vincent}, {V{\"o}lk}, \&
  {Wagner}}]{2005A&A...441..465A}
{Aharonian}, F., {et~al.} (H.E.S.S. Collab.) 2005,
  \aap, 441, 465

\bibitem[{{Aharonian} {et~al.}(2006){Aharonian}, {Akhperjanian}, {Bazer-Bachi},
  {Beilicke}, {Benbow}, {Berge}, {Bernl{\"o}hr}, {Boisson}, {Bolz}, {Borrel}, {Braun}, {Brown}, {B{\"u}hler}, {B{\"u}sching}, {Carrigan}, {Chadwick}, {Chounet}, {Coignet}, {Cornils}, {Costamante}, {Degrange}, {Dickinson}, {Djannati-Ata{\"i}}, {Drury}, {Dubus}, {Egberts},{Emmanoulopoulos}, {Espigat}, {Feinstein}, {Ferrero},{Fiasson}, {Fontaine}, {Funk}, {Funk},{F{\"u}{\ss}ling}, {Gallant}, {Giebels}, {Glicenstein},{Goret}, {Hadjichristidis}, {Hauser}, {Hauser},{Heinzelmann}, {Henri}, {Hermann}, {Hinton}, {Hoffmann}, {Hofmann}, {Holleran}, {Hoppe}, {Horns}, {Jacholkowska}, {de~Jager}, {Kendziorra}, {Kerschhaggl}, {Kh{\'e}lifi}, {Komin}, {Konopelko}, {Kosack}, {Lamanna}, {Latham}, {Le~Gallou}, {Lemi{\`e}re}, {Lemoine-Goumard}, {Lenain}, {Lohse}, {Martin}, {Martineau-Huynh}, {Marcowith}, {Masterson}, {Maurin}, {McComb}, {Moulin}, {de~Naurois}, {Nedbal}, {Nolan}, {Noutsos}, {Orford}, {Osborne}, {Ouchrif}, {Panter}, {Pelletier}, {Pita}, {P{\"u}hlhofer}, {Punch}, {Ranchon}, {Raubenheimer}, {Raue}, {Rayner}, {Reimer}, {Ripken}, {Rob}, {Rolland}, {Rosier-Lees}, {Rowell}, {Sahakian}, {Santangelo}, {Saug{\'e}}, {Schlenker}, {Schlickeiser}, {Schr{\"o}der}, {Schwanke}, {Schwarzburg}, {Schwemmer}, {Shalchi}, {Sol}, {Spangler}, {Spanier}, {Steenkamp}, {Stegmann}, {Superina}, {Tam}, {Tavernet}, {Terrier}, {Tluczykont}, {van~Eldik}, {Vasileiadis}, {Venter}, {Vialle}, {Vincent}, {V{\"o}lk}, {Wagner}, \& {Ward}}]{2006Sci...314.1424A}
{Aharonian}, F., {et~al.} (H.E.S.S. Collab.) 2006,
  Science, 314, 1424

\bibitem[{{Antonucci}(1993)}]{1993ARA&A..31..473A}
{Antonucci}, R. 1993, \araa, 31, 473

\bibitem[{{Bai} \& {Lee}(2001)}]{2001ApJ...549L.173B}
{Bai}, J.~M. \& {Lee}, M.~G. 2001, \apjl, 549, L173

\bibitem[{{Beilicke} {et~al.}(2005){Beilicke}, {Benbow}, {Cornils},
  {Heinzelmann}, {Horns}, {Raue}, {Ripken}, {Tluczykont}, \& {for the
  H.~E.~S.~S.~Collaboration}}]{2005astro.ph..4395B}
{Beilicke}, M., {Benbow}, W., {Cornils}, R., {et~al.} 2005, ArXiv Astrophysics e-prints, [{\tt arXiv:astro-ph/0504395}]

\bibitem[{{Beilicke} {et~al.}(2004){Beilicke}, {G{\"o}tting}, \&
  {Tluczykont}}]{2004NewAR..48..407B}
{Beilicke}, M., {G{\"o}tting}, N., \& {Tluczykont}, M. 2004, New Astronomy
  Review, 48, 407

\bibitem[{{Bergh{\"o}fer} {et~al.}(2000){Bergh{\"o}fer}, {Bowyer}, \&
  {Korpela}}]{2000ApJ...535..615B}
{Bergh{\"o}fer}, T.~W., {Bowyer}, S., \& {Korpela}, E. 2000, \apj, 535, 615

\bibitem[{{Biretta} {et~al.}(2002){Biretta}, {Junor}, \&
  {Livio}}]{2002NewAR..46..239B}
{Biretta}, J.~A., {Junor}, W., \& {Livio}, M. 2002, New Astronomy Review, 46,
  239

\bibitem[{{Biretta} {et~al.}(1999){Biretta}, {Sparks}, \&
  {Macchetto}}]{1999ApJ...520..621B}
{Biretta}, J.~A., {Sparks}, W.~B., \& {Macchetto}, F. 1999, \apj, 520, 621

\bibitem[{{Biretta} {et~al.}(1991){Biretta}, {Stern}, \&
  {Harris}}]{1991AJ....101.1632B}
{Biretta}, J.~A., {Stern}, C.~P., \& {Harris}, D.~E. 1991, \aj, 101, 1632

\bibitem[{{Bloom} \& {Marscher}(1996)}]{1996ApJ...461..657B}
{Bloom}, S.~D. \& {Marscher}, A.~P. 1996, \apj, 461, 657

\bibitem[{{B{\"o}hringer} {et~al.}(2001){B{\"o}hringer}, {Belsole}, {Kennea},
  {Matsushita}, {Molendi}, {Worrall}, {Mushotzky}, {Ehle}, {Guainazzi},
  {Sakelliou}, {Stewart}, {Vestrand}, \& {Dos Santos}}]{2001A&A...365L.181B}
{B{\"o}hringer}, H., {Belsole}, E., {Kennea}, J., {et~al.} 2001, \aap, 365,
  L181

\bibitem[{{Burns} {et~al.}(1983){Burns}, {Feigelson}, \&
  {Schreier}}]{1983ApJ...273..128B}
{Burns}, J.~O., {Feigelson}, E.~D., \& {Schreier}, E.~J. 1983, \apj, 273, 128

\bibitem[{{Capetti} {et~al.}(2000){Capetti}, {Trussoni}, {Celotti}, {Feretti},
  \& {Chiaberge}}]{2000MNRAS.318..493C}
{Capetti}, A., {Trussoni}, E., {Celotti}, A., {Feretti}, L., \& {Chiaberge}, M.
  2000, \mnras, 318, 493

\bibitem[{{Celotti} {et~al.}(1993){Celotti}, {Maraschi}, {Ghisellini},
  {Caccianiga}, \& {Maccacaro}}]{1993ApJ...416..118C}
{Celotti}, A., {Maraschi}, L., {Ghisellini}, G., {Caccianiga}, A., \&
  {Maccacaro}, T. 1993, \apj, 416, 118

\bibitem[{{Chiaberge} {et~al.}(2001){Chiaberge}, {Capetti}, \&
  {Celotti}}]{2001MNRAS.324L..33C}
{Chiaberge}, M., {Capetti}, A., \& {Celotti}, A. 2001, \mnras, 324, L33

\bibitem[{{Chiaberge} \& {Ghisellini}(1999)}]{1999MNRAS.306..551C}
{Chiaberge}, M. \& {Ghisellini}, G. 1999, \mnras, 306, 551

\bibitem[{{Collmar} {et~al.}(2000){Collmar}, {Reimer}, {Bennett}, {Bloemen},
  {Hermsen}, {Lichti}, {Ryan}, {Sch{\"o}nfelder}, {Steinle}, {Williams}, \&
  {B{\"o}ttcher}}]{2000A&A...354..513C}
{Collmar}, W., {Reimer}, O., {Bennett}, K., {et~al.} 2000, \aap, 354, 513

\bibitem[{{Dermer} \& {Gehrels}(1995)}]{1995ApJ...447..103D}
{Dermer}, C.~D. \& {Gehrels}, N. 1995, \apj, 447, 103

\bibitem[{{Evans} {et~al.}(2004){Evans}, {Kraft}, {Worrall}, {Hardcastle},
  {Jones}, {Forman}, \& {Murray}}]{2004ApJ...612..786E}
{Evans}, D.~A., {Kraft}, R.~P., {Worrall}, D.~M., {et~al.} 2004, \apj, 612, 786

\bibitem[{{Fioc} \& {Rocca-Volmerange}(1997)}]{1997A&A...326..950F}
{Fioc}, M. \& {Rocca-Volmerange}, B. 1997, \aap, 326, 950

\bibitem[{{Foschini} {et~al.}(2006){Foschini}, {Ghisellini}, {Raiteri},
  {Tavecchio}, {Villata}, {Maraschi}, {Pian}, {Tagliaferri}, {di Cocco}, \&
  {Malaguti}}]{2006A&A...453..829F}
{Foschini}, L., {Ghisellini}, G., {Raiteri}, C.~M., {et~al.} 2006, \aap, 453,
  829

\bibitem[{{Fossati} {et~al.}(1998){Fossati}, {Maraschi}, {Celotti}, {Comastri},
  \& {Ghisellini}}]{1998MNRAS.299..433F}
{Fossati}, G., {Maraschi}, L., {Celotti}, A., {Comastri}, A., \& {Ghisellini},
  G. 1998, \mnras, 299, 433

\bibitem[{{Georganopoulos} {et~al.}(2005){Georganopoulos}, {Perlman}, \&
  {Kazanas}}]{2005ApJ...634L..33G}
{Georganopoulos}, M., {Perlman}, E.~S., \& {Kazanas}, D. 2005, \apjl, 634, L33

\bibitem[{{Georganopoulos} {et~al.}(2006){Georganopoulos}, {Perlman},
  {Kazanas}, \& {McEnery}}]{2006ApJ...653L...5G}
{Georganopoulos}, M., {Perlman}, E.~S., {Kazanas}, D., \& {McEnery}, J. 2006,
  \apjl, 653, L5

\bibitem[{{Ghisellini} {et~al.}(1998){Ghisellini}, {Celotti}, {Fossati},
  {Maraschi}, \& {Comastri}}]{1998MNRAS.301..451G}
{Ghisellini}, G., {Celotti}, A., {Fossati}, G., {Maraschi}, L., \& {Comastri},
  A. 1998, \mnras, 301, 451

\bibitem[{{Giommi} {et~al.}(2002){Giommi}, {Capalbi}, {Fiocchi}, {Memola},
  {Perri}, {Piranomonte}, {Rebecchi}, \& {Massaro}}]{2002babs.conf...63G}
{Giommi}, P., {Capalbi}, M., {Fiocchi}, M., {et~al.} 2002, in Blazar
  Astrophysics with BeppoSAX and Other Observatories, ed. P.~{Giommi},
  E.~{Massaro}, \& G.~{Palumbo}, 63--+

\bibitem[{{Gopal-Krishna} {et~al.}(2007){Gopal-Krishna}, {Dhurde}, {Sircar}, \&
  {Wiita}}]{2007MNRAS.377..446G}
{Gopal-Krishna}, {Dhurde}, S., {Sircar}, P., \& {Wiita}, P.~J. 2007, \mnras,
  377, 446

\bibitem[{{Gopal-Krishna} {et~al.}(2006){Gopal-Krishna}, {Wiita}, \&
  {Dhurde}}]{2006MNRAS.369.1287G}
{Gopal-Krishna}, {Wiita}, P.~J., \& {Dhurde}, S. 2006, \mnras, 369, 1287

\bibitem[{{Gould}(1979)}]{1979A&A....76..306G}
{Gould}, R.~J. 1979, \aap, 76, 306

\bibitem[{{Graham}(1978)}]{1978PASP...90..237G}
{Graham}, J.~A. 1978, \pasp, 90, 237

\bibitem[{{Grandi} {et~al.}(2006){Grandi}, {Malaguti}, \&
  {Fiocchi}}]{2006ApJ...642..113G}
{Grandi}, P., {Malaguti}, G., \& {Fiocchi}, M. 2006, \apj, 642, 113

\bibitem[{{Grandi} \& {Palumbo}(2007)}]{2007ApJ...659..235G}
{Grandi}, P. \& {Palumbo}, G.~G.~C. 2007, \apj, 659, 235

\bibitem[{{Hardcastle} {et~al.}(2003){Hardcastle}, {Worrall}, {Kraft},
  {Forman}, {Jones}, \& {Murray}}]{2003ApJ...593..169H}
{Hardcastle}, M.~J., {Worrall}, D.~M., {Kraft}, R.~P., {et~al.} 2003, \apj,
  593, 169

\bibitem[{{H{\"a}ring-Neumayer} {et~al.}(2006){H{\"a}ring-Neumayer},
  {Cappellari}, {Rix}, {Hartung}, {Prieto}, {Meisenheimer}, \&
  {Lenzen}}]{2006ApJ...643..226H}
{H{\"a}ring-Neumayer}, N., {Cappellari}, M., {Rix}, H.-W., {et~al.} 2006, \apj,
  643, 226

\bibitem[{{Hartman} {et~al.}(1999){Hartman}, {Bertsch}, {Bloom}, {Chen},
  {Deines-Jones}, {Esposito}, {Fichtel}, {Friedlander}, {Hunter}, {McDonald},
  {Sreekumar}, {Thompson}, {Jones}, {Lin}, {Michelson}, {Nolan}, {Tompkins},
  {Kanbach}, {Mayer-Hasselwander}, {M{\"u}cke}, {Pohl}, {Reimer}, {Kniffen},
  {Schneid}, {von Montigny}, {Mukherjee}, \& {Dingus}}]{1999ApJS..123...79H}
{Hartman}, R.~C., {Bertsch}, D.~L., {Bloom}, S.~D., {et~al.} 1999, \apjs, 123,
  79

\bibitem[{{Hawarden} {et~al.}(1993){Hawarden}, {Sandell}, {Matthews},
  {Friberg}, {Watt}, \& {Smith}}]{1993MNRAS.260..844H}
{Hawarden}, T.~G., {Sandell}, G., {Matthews}, H.~E., {et~al.} 1993, \mnras,
  260, 844

\bibitem[{{Henri} \& {Saug{\'e}}(2006)}]{2006ApJ...640..185H}
{Henri}, G. \& {Saug{\'e}}, L. 2006, \apj, 640, 185

\bibitem[{{Horiuchi} {et~al.}(2006){Horiuchi}, {Meier}, {Preston}, \&
  {Tingay}}]{2006PASJ...58..211H}
{Horiuchi}, S., {Meier}, D.~L., {Preston}, R.~A., \& {Tingay}, S.~J. 2006,
  \pasj, 58, 211

\bibitem[{{Inoue} \& {Takahara}(1996)}]{1996ApJ...463..555I}
{Inoue}, S. \& {Takahara}, F. 1996, \apj, 463, 555

\bibitem[{{Katarzy{\'n}ski} {et~al.}(2001){Katarzy{\'n}ski}, {Sol}, \&
  {Kus}}]{2001A&A...367..809K}
{Katarzy{\'n}ski}, K., {Sol}, H., \& {Kus}, A. 2001, \aap, 367, 809

\bibitem[{{Katarzy{\'n}ski} {et~al.}(2003){Katarzy{\'n}ski}, {Sol}, \&
  {Kus}}]{2003A&A...410..101K}
{Katarzy{\'n}ski}, K., {Sol}, H., \& {Kus}, A. 2003, \aap, 410, 101

\bibitem[{{Keel}(1985)}]{1985AJ.....90.2207K}
{Keel}, W.~C. 1985, \aj, 90, 2207

\bibitem[{{Kinzer} {et~al.}(1995){Kinzer}, {Johnson}, {Dermer}, {Kurfess},
  {Strickman}, {Grove}, {Kroeger}, {Grabelsky}, {Purcell}, {Ulmer}, {Jung}, \&
  {McNaron-Brown}}]{1995ApJ...449..105K}
{Kinzer}, R.~L., {Johnson}, W.~N., {Dermer}, C.~D., {et~al.} 1995, \apj, 449,
  105

\bibitem[{{Le Bohec} {et~al.}(2004){Le Bohec}, {Badran}, {Bond}, {Boyle},
  {Bradbury}, {Buckley}, {Carter-Lewis}, {Catanese}, {Celik}, {Cui}, {Daniel},
  {D'Vali}, {de la Calle Perez}, {Duke}, {Falcone}, {Fegan}, {Fegan}, {Finley},
  {Fortson}, {Gaidos}, {Gammell}, {Gibbs}, {Gillanders}, {Grube}, {Hall},
  {Hall}, {Hanna}, {Hillas}, {Holder}, {Horan}, {Jarvis}, {Jordan}, {Kenny},
  {Kertzman}, {Kieda}, {Kildea}, {Knapp}, {Kosack}, {Krawczynski}, {Krennrich},
  {Lang}, {Linton}, {Lloyd-Evans}, {Milovanovic}, {Moriarty}, {M{\"u}ller},
  {Nagai}, {Nolan}, {Ong}, {Pallassini}, {Petry}, {Power-Mooney}, {Quinn},
  {Quinn}, {Ragan}, {Rebillot}, {Reynolds}, {Rose}, {Schroedter}, {Sembroski},
  {Swordy}, {Syson}, {Vassiliev}, {Wakely}, {Walker}, {Weekes}, \&
  {Zweerink}}]{2004ApJ...610..156L}
{Le Bohec}, S., {Badran}, H.~M., {Bond}, I.~H., {et~al.} 2004, \apj, 610, 156

\bibitem[{{Macchetto} {et~al.}(1997){Macchetto}, {Marconi}, {Axon}, {Capetti},
  {Sparks}, \& {Crane}}]{1997ApJ...489..579M}
{Macchetto}, F., {Marconi}, A., {Axon}, D.~J., {et~al.} 1997, \apj, 489, 579

\bibitem[{{Marconi} {et~al.}(2006){Marconi}, {Pastorini}, {Pacini}, {Axon},
  {Capetti}, {Macchetto}, {Koekemoer}, \& {Schreier}}]{2006A&A...448..921M}
{Marconi}, A., {Pastorini}, G., {Pacini}, F., {et~al.} 2006, \aap, 448, 921

\bibitem[{{Marconi} {et~al.}(2000){Marconi}, {Schreier}, {Koekemoer},
  {Capetti}, {Axon}, {Macchetto}, \& {Caon}}]{2000ApJ...528..276M}
{Marconi}, A., {Schreier}, E.~J., {Koekemoer}, A., {et~al.} 2000, \apj, 528,
  276

\bibitem[{{McKinney}(2006)}]{2006MNRAS.368.1561M}
{McKinney}, J.~C. 2006, \mnras, 368, 1561

\bibitem[{{Melia} {et~al.}(2002){Melia}, {Liu}, \&
  {Fatuzzo}}]{2002ApJ...567..811M}
{Melia}, F., {Liu}, S., \& {Fatuzzo}, M. 2002, \apj, 567, 811

\bibitem[{{Mirabel} {et~al.}(1999){Mirabel}, {Laurent}, {Sanders}, {Sauvage},
  {Tagger}, {Charmandaris}, {Vigroux}, {Gallais}, {Cesarsky}, \&
  {Block}}]{1999A&A...341..667M}
{Mirabel}, I.~F., {Laurent}, O., {Sanders}, D.~B., {et~al.} 1999, \aap, 341,
  667

\bibitem[{{Neronov} \& {Aharonian}(2007)}]{2007arXiv0704.3282N} {Neronov}, A., \& {Aharonian}, F.\ 2007, ArXiv e-prints, 704, [{\tt arXiv:0704.3282}]

\bibitem[{{Padovani}(2007)}]{2007Ap&SS.tmp..190P}
{Padovani}, P. 2007, \apss, 190

\bibitem[{{Paltani} \& {T{\"u}rler}(2005)}]{2005A&A...435..811P}
{Paltani}, S. \& {T{\"u}rler}, M. 2005, \aap, 435, 811

\bibitem[{{Perlman} {et~al.}(2003){Perlman}, {Harris}, {Biretta}, {Sparks}, \&
  {Macchetto}}]{2003ApJ...599L..65P}
{Perlman}, E.~S., {Harris}, D.~E., {Biretta}, J.~A., {Sparks}, W.~B., \&
  {Macchetto}, F.~D. 2003, \apjl, 599, L65

\bibitem[{{Perlman} {et~al.}(2001){Perlman}, {Sparks}, {Radomski}, {Packham},
  {Fisher}, {Pi{\~n}a}, \& {Biretta}}]{2001ApJ...561L..51P}
{Perlman}, E.~S., {Sparks}, W.~B., {Radomski}, J., {et~al.} 2001, \apjl, 561,
  L51

\bibitem[{{Perlman} \& {Wilson}(2005)}]{2005ApJ...627..140P}
{Perlman}, E.~S. \& {Wilson}, A.~S. 2005, \apj, 627, 140

\bibitem[{{Pian} {et~al.}(1996){Pian}, {Falomo}, {Ghisellini}, {Maraschi},
  {Sambruna}, {Scarpa}, \& {Treves}}]{1996ApJ...459..169P}
{Pian}, E., {Falomo}, R., {Ghisellini}, G., {et~al.} 1996, \apj, 459, 169

\bibitem[{{Reynolds} {et~al.}(1996){Reynolds}, {Fabian}, {Celotti}, \&
  {Rees}}]{1996MNRAS.283..873R}
{Reynolds}, C.~S., {Fabian}, A.~C., {Celotti}, A., \& {Rees}, M.~J. 1996,
  \mnras, 283, 873

\bibitem[{{Roberts} {et~al.}(1998){Roberts}, {Dazeley}, {Edwards}, {Hara},
  {Hayami}, {Holder}, {Kakimoto}, {Kamei}, {Kawachi}, {Kifune}, {Kita},
  {Konishi}, {Masaike}, {Matsubara}, {Matsuoka}, {Mizumoto}, {Mori},
  {Muraishi}, {Muraki}, {Nishijima}, {Oda}, {Ogio}, {Patterson}, {Rowell},
  {Sako}, {Sakurazawa}, {Susukita}, {Suzuki}, {Suzuki}, {Tamura}, {Tanimori},
  {Thornton}, {Yanagita}, {Yoshida}, \& {Yoshikoshi}}]{1998A&A...337...25R}
{Roberts}, M.~D., {et~al.} (CANGAROO Collab.) 1998, \aap, 337,
  25

\bibitem[{{Rothschild} {et~al.}(2006){Rothschild}, {Wilms}, {Tomsick},
  {Staubert}, {Benlloch}, {Collmar}, {Madejski}, {Deluit}, \&
  {Khandrika}}]{2006ApJ...641..801R}
{Rothschild}, R.~E., {Wilms}, J., {Tomsick}, J., {et~al.} 2006, \apj, 641, 801

\bibitem[{{Schmidt}(1963)}]{1963Natur.197.1040S}
{Schmidt}, M. 1963, \nat, 197, 1040

\bibitem[{{Smith} {et~al.}(2000){Smith}, {Lucey}, {Hudson}, {Schlegel}, \&
  {Davies}}]{2000MNRAS.313..469S}
{Smith}, R.~J., {Lucey}, J.~R., {Hudson}, M.~J., {Schlegel}, D.~J., \&
  {Davies}, R.~L. 2000, \mnras, 313, 469

\bibitem[{{Sparks} {et~al.}(1996){Sparks}, {Biretta}, \&
  {Macchetto}}]{1996ApJ...473..254S}
{Sparks}, W.~B., {Biretta}, J.~A., \& {Macchetto}, F. 1996, \apj, 473, 254

\bibitem[{{Sreekumar} {et~al.}(1994){Sreekumar}, {Bertsch}, {Dingus},
  {Esposito}, {Fichtel}, {Hartman}, {Hunter}, {Kanbach}, {Kniffen}, {Lin},
  {Mattox}, {Mayer-Hasselwander}, {Michelson}, {von Montigny}, {Nolan},
  {Schneid}, \& {Thompson}}]{1994ApJ...426..105S}
{Sreekumar}, P., {Bertsch}, D.~L., {Dingus}, B.~L., {et~al.} 1994, \apj, 426,
  105

\bibitem[{{Stawarz} {et~al.}(2006){Stawarz}, {Aharonian}, {Wagner}, \&
  {Ostrowski}}]{2006MNRAS.371.1705S}
{Stawarz}, {\L}., {Aharonian}, F., {Wagner}, S., \& {Ostrowski}, M. 2006,
  \mnras, 371, 1705

\bibitem[{{Stawarz} {et~al.}(2003){Stawarz}, {Sikora}, \&
  {Ostrowski}}]{2003ApJ...597..186S}
{Stawarz}, {\L}., {Sikora}, M., \& {Ostrowski}, M. 2003, \apj, 597, 186

\bibitem[{{Stecker} {et~al.}(2006){Stecker}, {Malkan}, \&
  {Scully}}]{2006ApJ...648..774S}
{Stecker}, F.~W., {Malkan}, M.~A., \& {Scully}, S.~T. 2006, \apj, 648, 774

\bibitem[{{Steinle} {et~al.}(1998){Steinle}, {Bennett}, {Bloemen}, {Collmar},
  {Diehl}, {Hermsen}, {Lichti}, {Morris}, {Schonfelder}, {Strong}, \&
  {Williams}}]{1998A&A...330...97S}
{Steinle}, H., {Bennett}, K., {Bloemen}, H., {et~al.} 1998, \aap, 330, 97

\bibitem[{{Strauss} {et~al.}(1992){Strauss}, {Huchra}, {Davis}, {Yahil},
  {Fisher}, \& {Tonry}}]{1992ApJS...83...29S}
{Strauss}, M.~A., {Huchra}, J.~P., {Davis}, M., {et~al.} 1992, \apjs, 83, 29

\bibitem[{{Thompson} {et~al.}(1995){Thompson}, {Bertsch}, {Dingus}, {Esposito},
  {Etienne}, {Fichtel}, {Friedlander}, {Hartman}, {Hunter}, {Kendig}, {Mattox},
  {McDonald}, {von Montigny}, {Mukherjee}, {Ramanamurthy}, {Sreekumar},
  {Fierro}, {Lin}, {Michelson}, {Nolan}, {Shriver}, {Willis}, {Kanbach},
  {Mayer-Hasselwander}, {Merck}, {Radecke}, {Kniffen}, \&
  {Schneid}}]{1995ApJS..101..259T}
{Thompson}, D.~J., {Bertsch}, D.~L., {Dingus}, B.~L., {et~al.} 1995, \apjs,
  101, 259

\bibitem[{{Tingay} \& {Edwards}(2002)}]{2002AJ....124..652T}
{Tingay}, S.~J. \& {Edwards}, P.~G. 2002, \aj, 124, 652

\bibitem[{{Tingay} {et~al.}(1998){Tingay}, {Jauncey}, {Reynolds}, {Tzioumis},
  {King}, {Preston}, {Jones}, {Murphy}, {Meier}, {van Ommen}, {McCulloch},
  {Ellingsen}, {Costa}, {Edwards}, {Lovell}, {Nicolson}, {Quick}, {Kemball},
  {Migenes}, {Harbison}, {Jones}, {White}, {Gough}, {Ferris}, {Sinclair}, \&
  {Clay}}]{1998AJ....115..960T}
{Tingay}, S.~J., {Jauncey}, D.~L., {Reynolds}, J.~E., {et~al.} 1998, \aj, 115,
  960

\bibitem[{{Tornikoski} {et~al.}(2002){Tornikoski}, {L{\"a}hteenm{\"a}ki},
  {Lainela}, \& {Valtaoja}}]{2002ApJ...579..136T}
{Tornikoski}, M., {L{\"a}hteenm{\"a}ki}, A., {Lainela}, M., \& {Valtaoja}, E.
  2002, \apj, 579, 136

\bibitem[{{T{\"u}rler} {et~al.}(2006){T{\"u}rler}, {Chernyakova},
  {Courvoisier}, {Foellmi}, {Aller}, {Aller}, {Kraus}, {Krichbaum},
  {L{\"a}hteenm{\"a}ki}, {Marscher}, {McHardy}, {O'Brien}, {Page}, {Popescu},
  {Robson}, {Tornikoski}, \& {Ungerechts}}]{2006A&A...451L...1T}
{T{\"u}rler}, M., {Chernyakova}, M., {Courvoisier}, T.~J.-L., {et~al.} 2006,
  \aap, 451, L1

\bibitem[{{T{\"u}rler} {et~al.}(1999){T{\"u}rler}, {Paltani}, {Courvoisier},
  {Aller}, {Aller}, {Blecha}, {Bouchet}, {Lainela}, {McHardy}, {Robson},
  {Stevens}, {Ter{\"a}sranta}, {Tornikoski}, {Ulrich}, {Waltman}, {Wamsteker},
  \& {Wright}}]{1999A&AS..134...89T}
{T{\"u}rler}, M., {Paltani}, S., {Courvoisier}, T.~J.-L., {et~al.} 1999, \aaps,
  134, 89

\bibitem[{{Ulrich}(1981)}]{1981A&A...103L...1U}
{Ulrich}, M.~H. 1981, \aap, 103, L1+

\bibitem[{{Unwin} {et~al.}(1985){Unwin}, {Cohen}, {Biretta}, {Pearson},
  {Seielstad}, {Walker}, {Simon}, \& {Linfield}}]{1985ApJ...289..109U}
{Unwin}, S.~C., {Cohen}, M.~H., {Biretta}, J.~A., {et~al.} 1985, \apj, 289, 109

\bibitem[{{Urry} \& {Padovani}(1995)}]{1995PASP..107..803U}
{Urry}, C.~M. \& {Padovani}, P. 1995, \pasp, 107, 803

\bibitem[{{Vlahakis} \& {K{\"o}nigl}(2004)}]{2004ApJ...605..656V}
{Vlahakis}, N. \& {K{\"o}nigl}, A. 2004, \apj, 605, 656

\bibitem[{{Wiita}(2006)}]{2006astro.ph..3728W} {Wiita}, P.~J. 2006, ArXiv 
Astrophysics e-prints, [{\tt arXiv:astro-ph/0603728}]

\bibitem[{{Wilson} \& {Yang}(2002)}]{2002ApJ...568..133W}
{Wilson}, A.~S. \& {Yang}, Y. 2002, \apj, 568, 133

\bibitem[{{Woo} {et~al.}(2005){Woo}, {Urry}, {van der Marel}, {Lira}, \&
  {Maza}}]{2005ApJ...631..762W}
{Woo}, J.-H., {Urry}, C.~M., {van der Marel}, R.~P., {Lira}, P., \& {Maza}, J.
  2005, \apj, 631, 762

\end{thebibliography}
     
\end{document}